# Efficient Electrical Control of Thin-Film Black Phosphorus Bandgap

Bingchen Deng[1+], Vy Tran[2+], Hao Jiang[3], Cheng Li[1], Yujun (Terry) Xie[4], Qiushi Guo[1], Xiaomu Wang[1], He Tian[5], Han Wang[5], Judy J. Cha[4], Qiangfei Xia[3], Li Yang[2*], and Fengnian Xia[1*]

[1]Department of Electrical Engineering, Yale University, New Haven, Connecticut 06511, USA

[2]Department of Physics, Washington University, St Louis, Missouri 63130, USA

[3]Department of Electrical & Computer Engineering, University of Massachusetts, Amherst, Massachusetts 01003, USA

[4]Department of Mechanical Engineering and Materials Science, Yale University, New Haven, Connecticut 06511, USA

[5]Ming Hsieh Department of Electrical Engineering, University of Southern California, Los Angeles, California 90089, USA

Recently rediscovered black phosphorus is a layered semiconductor with promising electronic and photonic properties[1-10]. Dynamic control of its bandgap can enable novel device applications and allow for the exploration of new physical phenomena. However, theoretical investigations[11-14] and photoemission spectroscopy experiments performed on doped black phosphorus through potassium adsorption[15] indicate that in its few-layer form, an exceedingly large electric field in the order of several volts per nanometer is required to effectively tune its bandgap, making the direct electrical control unfeasible. Here we demonstrate the tuning of bandgap in intrinsic black phosphorus using an electric field directly and reveal the unique thickness-dependent bandgap tuning properties, arising from the strong interlayer electronic-state coupling. Furthermore, leveraging a 10-nm-thick black phosphorus in which the field-induced potential difference across the film dominates over the interlayer coupling, we continuously tune its bandgap from ~300 to below 50 milli-electron volts, using a moderate displacement field up to 1.1 volts per nanometer. Such dynamic tuning of bandgap may not only extend the operational wavelength range of tunable black phosphorus photonic devices, but also pave the way for the investigation of electrically tunable topological insulators and topological nodal semimetals[14-16].

[+]Contributed Equally

*lyang@physics.wustl.edu; *fengnian.xia@yale.edu;

Bandgap is a fundamental material parameter which governs the transport and light-matter interaction properties. Black phosphorus (BP) lately emerged as a promising layered material which bridges the gap between the gapless graphene[17, 18] and semiconducting transition metal dichalcogenides (TMDCs)[19] such as molybdenum disulfide ($MoS_2$) with a relatively large bandgap of around 2 electron volts (eV). Due to the strong electronic state coupling among layers in BP, its direct bandgap varies significantly from around 0.3 eV in bulk to 2 eV in monolayer form[9]. On the other hand, an approach which can dynamically tune the bandgap of BP to well below 0.3 eV is of great scientific and technological importance. In fact, the feasibility of bandgap tuning in few-layer BP using an electric field has been explored theoretically[11-14]. Recently, photoemission spectroscopy experiments in doped few-layer BP through the adsorption of potassium atoms show that indeed its bandgap can be tuned and even completely closed[15]. However, the estimated electric field induced by potassium adsorption in few-layer BP to widely tune its bandgap is in the order of several volts per nanometer. Such a large electric field prevents the realization of bandgap tunable electronic and photonic devices. Moreover, heavily doped material is usually not ideal for device applications. As a result, it is highly attractive to realize a widely tunable bandgap in intrinsic BP with a moderate field accessible with regular dielectrics.

In this work, we demonstrate the bandgap tuning in BP using an electric field directly and discover the unique thickness-dependent bandgap tuning properties, guided by the first-principles calculations. In a 4-nm-thick intrinsic BP film, the bandgap tuning is limited to around 75 meV and shows strong non-linear dependence on the biasing field. This peculiar field dependence is due to the strong interlayer electronic-state coupling and is captured by density functional theory (DFT) calculations very well. On the contrary, in a 10-nm-thick intrinsic BP in which the film is thick enough such that the field-induced potential difference can overcome the interlayer coupling, we demonstrate the efficient, continuous tuning of its bandgap of from ~300 to below 50 meV with an external displacement field up to 1.1 volt per nanometer. We further show that in this case, the additional dielectric screening effect beyond a simple DFT prediction due to the significant bandgap shrinkage has to be taken into account to describe the experimental results.

We utilize dual-gate BP transistors as shown in Fig. 1a for the bandgap measurement under a bias field. Here, the thin-film BP channel is sandwiched between the back (90-nm-thick silicon oxide) and top (24-nm-thick aluminum oxide) gate dielectrics. The source and drain leads are made from chromium/gold (3/30 nm). The top gate is made from titanium/platinum (1/10 nm) and the silicon substrate is used as the back gate. The bandgap is determined through the BP conductance at the charge neutrality point under a vertical biasing field. In order to eliminate the impact of metal/BP contact on the conductance measurement, a 4-probe scheme is adopted, as shown in Fig. 1a. In our measurements, the source is grounded and the drain bias $V_D$ is kept at 100 mV. From the channel current and the voltage difference between probes 1 and 2, we determine the conductance of the BP film. The detailed device fabrication and measurement procedures are discussed in the Methods section. In Fig. 1b, the measured conductance $G_{12}$ of a 4-nm-thick BP flake between probes 1 and 2 (see inset) at room temperature are plotted as a function of the back gate bias, $V_{BG}$, while the top gate bias, $V_{TG}$, is set to be 0 volt. Here the BP thickness is determined using the atomic force microscopy (AFM) together with the cross-section high- resolution transmission electron microscopy (HRTEM), as shown in the Supplementary Information I. In short, the thickness of BP measured using AFM is usually 2 to 3 nanometers larger than the intrinsic BP thickness due to the top surface oxidation during the device fabrication. The conductance is plotted in both linear (left axis) and logarithmic (right axis) scales. Because the bandgap of BP is moderate, the conductance in insulating state at room temperature can still be measured using 4-probe scheme, as shown by the logarithmic plot (right axis, Fig. 1b).

We further performed conductance measurements in such dual-gated BP transistors by sweeping the top gate bias at different static back gate biases. The measurement results for a BP transistor with a channel thickness of around 4-nm are summarized in Fig. 2a. As shown in the inset of Fig. 2b, the top gate bias at which the conductance is minimized ($V_{TM}$) scales almost linearly with the back gate bias ($V_{BG}$). This observation indicates that the top gate bias with opposite sign can effectively compensate the doping introduced by the back gate at the $V_{TM}$, leading to an insulating state at the charge neutrality point. Since the source is grounded and a very small drain bias of 100 mV is applied, the displacement fields in the back ($D_B$) and top ($D_T$) gate dielectric are $D_B = \varepsilon_B V_{BG}/d_B$ and $D_T = -\varepsilon_T V_{TG}/d_T$, respectively, where $\varepsilon_B$ (~3.9) is the relative permittivity of $SiO_2$, $d_B$ (90 nm) is the thickness of $SiO_2$, $\varepsilon_T$ is the overall relative permittivity of top gate dielectric, and $d_T$ (26 nm) is the total thickness of top gate dielectric. Here displacement is defined as positive when it points from back towards the top gates. The $d_T$ includes the 2-nm-thick phosphorus oxide layer and 24-nm-thick $Al_2O_3$. From the slope of the curve in the inset of Fig. 2b, $\varepsilon_T$ is determined to be ~5.6. At the charge-neutrality condition ($V_{TG} = V_{TM}$), $D_B = D_T = D$, where $D$ is the displacement field in black phosphorus channel, as shown in Fig. 1a. We assume that at the charge-neutrality condition, the free carrier concentration in BP is minimized such that $D$ is approximately uniform across the entire channel.

In Fig. 2b, the minimal conductance at the charge-neutrality point is plotted as a function of the displacement field $D$ in BP. It increases by ~5 times at maximum bias field, clearly indicating a reduction of the energy gap. We quantify this bandgap reduction in the following. The minimum conductivity $\sigma_m$ at the charge neutrality can be calculated using

$$\sigma_m = qn_i(\mu_e + \mu_h) \qquad (1)$$

where $q$ is the elementary electron charge, $n_i$ is the intrinsic, thermally-excited carrier density for both electrons and holes at charge neutrality, and $\mu_e$ and $\mu_h$ are the mobility for electrons and holes, respectively. If the bandgap reduction is much smaller than the bandgap itself, we can assume that the band structure does not change and hence the carrier mobility remains unchanged. In this case, from the minimum conductivity variation, we can estimate the bandgap reduction since $n_i$ is proportional to $e^{-\frac{E_g}{2k_BT}}$, where $E_g$ is the bandgap, $k_B$ is the Boltzmann constant and $T$ is temperature (295 Kelvin)[20]. The red dots in Fig. 2c indicate the estimated bandgap shrinkage as a function of the externally applied displacement field, which does not exceed 75 meV and shows a strong non-linear dependence. Indeed, the reduction is much smaller than the quasi-particle bandgap of ~4-nm-thick black phosphorus (~ 7 layers), which should be around 450 meV[9].

We have performed theoretical calculations using density functional theory (DFT) to explore the bandgap tuning effect in BP thin film. Based on the DFT results, we have further developed a tight-binding model (see details in the Supplementary Information II), in which the interlayer coupling ($\delta$) is introduced in the off-diagonal terms of the Hamiltonian and the screened field induced potential difference ($\Delta$) between layers is placed in the diagonal terms. This model can reliably reproduce DFT results for few-layer BP and enable us to efficiently calculate much thicker BP that is beyond the DFT capability. As shown in Fig. 2c, our experimental results measured on a 4-nm-thick BP sample (~7 layers nominally) fall in between the theoretical 5- and 6-layer results. Most importantly, the theoretical results capture the non-linear character of the bandgap evolution as the gating field increases. In short, the bandgap tuning effect arises from the overall potential difference induced by the external gating field. However, for ultra-thin (few-layer) BP films, the potential drop is less than the interlayer coupling (off-diagonal term $\delta$) under low bias. Thus the bandgap tuning exhibits a very strong non-linear dependence on the external bias field and the tunable range is rather limited with a moderate displacement field up to ~1 V/nm. On the contrary, our model predicts that, when the thickness is large enough, the total potential difference across the entire BP thin-film dominates, leading to more effective and almost linear dependence of the tuning with the applied bias field as shown in Fig. 2c for thicker BP (16 to 19 layers). Although in general for thicker material the bandgap tuning effect should be larger because the Stark coefficient scales with thickness[14, 15], here we reveal the distinctively different tuning properties of BP films whose thicknesses fall into two different categories as shown in Fig. 2c.

Therefore, in order to extend the bandgap tuning range, we further fabricated the dual-gate transistors with BP channel thickness of ~10 nm. The conductance measurement results similar to those in 4-nm-thick transistors are plotted in Fig. 3a. The minimum conductance at the charge-neutrality point as a function of the external displacement field is plotted in Fig. 3b. In this transistor, the minimum conductance at the charge-neutrality increases by around 40 times as the external bias is maximized, indicating a bandgap reduction significantly greater than 75 meV. In this case, direct extraction of the bandgap variation from the minimum conductance at different biasing field may not be optimal since the bandstructure itself can vary as the biasing changes the bandgap significantly. In this transistor, we measured the temperature dependence of the minimum conductance to determine the energy gap. Using this method, we not only eliminate the effect of bandstructure change on bandgap determination, but also measure the bandgap directly. The measurement results for the back gate biases of -15, 0 and 15 V are shown in Fig. 4a. In Fig. 4b, similar curves for the back gate biases of -25 and 25 V are shown. The detailed bandgap extraction procedures and the typical fitting curves are presented in the Supplementary

Information III. Below 120 K, the BP film at zero bias becomes highly insulating and it is not feasible to measure its conductance. The minimum conductance at zero bias varies by around 4 orders of magnitude from 120 to 295 Kelvin as shown in Fig. 4a, from which a bandgap of 295 meV is extracted. This value is very close to the well-established bulk BP bandgap of ~300 meV, indicating the accuracy of the approach used here. When the applied displacement field increases from 0 to 1.1 V/nm ($V_{BG}$ positive and $V_{TG}$ negative), the bandgap is measured to decrease from 295 to 40 meV continuously, as shown in Fig. 4c. Reversing the direction of the displacement field leads to similar reduction of BP bandgap. When the temperature changes, the top gate bias at which the minimum conductance occurs, $V_{TM}$, varies slightly (< 0.5 V). We hence introduced an error bar to account for this uncertainty in external biasing displacement field in Fig. 4c.

In Fig. 4c, we also plot the calculated bandgap tuning results using tight-binding model built on DFT results for BP thickness from 9 to 11 nm (~17 to 21 layers), as shown by the dashed lines. It is clear that DFT predicts a faster bandgap tuning rate as a function of the biasing field. This discrepancy is not surprising because DFT is known for its deficiency to capture screening effects between electrons and subsequent dielectric function. Unlike the case of 4-nm samples, in which the bandgap variation (~75 meV) is rather smaller compared to the bandgap itself (~ 450 meV), the widely changed bandgap in the 10-nm BP makes it necessary to include the variation of dielectric screening beyond DFT. Using the random-phase approximation[21], in which the dominant contribution to the dielectric constant is from the interband transition around the bandgap, we introduced a gap dependent dielectric constant $\varepsilon(E_g) = \varepsilon_0 {E_{g0}}/{E_g}$ in our tight-binding model, where $\varepsilon_0$ is the dielectric constant of pristine BP, $E_{g0}$ is the bandgap of pristine BP (~ 300 meV in 10-nm-thick film), and $E_g$ is the bandgap under bias, which can be obtained self-consistently. This model reflects the enhanced dielectric screening when the bandgap is reduced. We have self-consistently calculated bandgaps for BP with thickness from 17 to 21 layers and the results are also plotted in Fig. 4c. The detailed calculation procedures are presented in Supplementary Information II. The measured bandgap of 10-nm-thick BP under bias in general agrees well with the self-consistent tight-binding model taking into account the additional dielectric screening. At vertical displacement field above 0.8 V/nm, it seems that the theoretical results underestimate the bandgap reduction, probably due to the oversimplified model for the dielectric constant.

In principle the bandgap tuning effect can be even larger in thicker black phosphorus thin-films. However, the physical picture above is only valid when the free carrier screening length is longer or comparable to the thickness of BP thin film. In this case, the top and back biases at opposite polarities can effectively compensate the free carriers introduced by each other, leading to an insulating state. In the Supplementary Information IV, we plot the measured conductance of a 23-nm-thick BP film using the 4-probe scheme. It is clear that the doping compensation picture described above does not apply to such thick BP films. As shown in the Supplementary Information IV, there the top and back gates largely modify the conductance of the top and bottom BP channels, respectively. Therefore, the internal electric field is no longer uniform; the field-induced electrons and holes near the surfaces[22] will screen the applied electric field. The linear relationship between $V_{TM}$ and $V_{BG}$ no longer exists and the thick BP film can no longer be regarded as a material system with a universal bandgap.

We want to emphasize that this work represents the demonstration of bandgap tuning in thin-film layered materials whose un-tuned properties are close to the bulk limit. This is particularly important for future device applications, since thin-film materials interact with light more strongly compared with mono- or few-layer materials and the carrier transport in thin-film is much less susceptible to environment. Moreover, tuning of bandgap in thin-film BP exhibits distinctively different properties compared with widely explored transition metal dichalcogenide (TMD) bilayers[23-25]. In bilayer TMDs such as $MoS_2$, due to the weak interlayer electronic state coupling, the bandgap can be tuned effectively and linearly with biasing field since the interlayer potential difference induced by the external field dominates even in bilayer TMDs[25]. However, bilayer $MoS_2$ has an indirect bandgap. Vertical electrical field changes the indirect bandgap while the direct optical transitions are hardly affected by the tuning[25]. On the contrary, black phosphorus is a direct bandgap material regardless of its layer number or applied field. Moreover, in black phosphorus the wave functions of the conduction band minimum (CBM) and valence band maximum (VBM) of thin-film BP are expected to exhibit substantial overlap even under bias due to the strong interlayer electronic state coupling. This character is of particular importance for optical properties, because the overlap gives rise to significant dipole oscillator strength, implying that the optical emission/absorption at the direct bandgap edge can be tuned by the gating-field approach.

In fact, even before the exploration of bilayer TMDs, bandgap opening in bilayer graphene using an vertical electric field has been extensively studied[26-34]. Here the tuning of the bandgap of BP also has distinctively different properties if compared with bilayer graphene. First, it is not feasible to realize bandgap tuning in thin-film graphite due to strong free carrier screening effect, arising from its metallic nature. Moreover, in general experimentally measured transport bandgaps in biased bilayer graphene are much less than the theoretical predictions and the values determined by optical experiments[30-34]. This is mainly because the spatial fluctuations of the gating field due to the interfacial charges and charges trapped in gate dielectrics lead to some ungated or weakly gated areas[34]. Although they probably represent only a small fraction of the bilayer sample area, these ungated or weakly gated areas are close to metallic and have large conductance. As a result, carrier transport is significantly affected by these highly conductive "leakage" paths[34], leading to discrepancy in theory and experiments. In BP thin film, although it is highly likely that there are also such ungated or weakly gated areas, those areas are semiconducting and the transport properties under the bias are dominated by the efficiently gated areas in which the bandgap is significantly reduced. As a result, our transport measurement results agree with theoretical values well.

In summary, we demonstrate the unique bandgap tuning properties of thin film BP and further show that the bandgap of a 10-nm-thick BP can be continuously tuned from ~300 meV to below 50 meV, with a moderate field readily achieved using regular dielectrics. This demonstration may significantly extend the operational wavelength range of BP optoelectronic devices, making it a technologically relevant material for tunable infrared applications beyond the cut-off wavelength of pristine BP (4 μm). The concept of bandgap tuning may also be leveraged to construct novel electronic devices. Interestingly, this demonstration may further enable the exploration of electrically tunable, BP based topological insulators and topological nodal semimetals.

## Methods

### Fabrication of dual-gated black phosphorus transistors

The fabrication of the devices started with the exfoliation of BP thin films from bulk crystals onto 90 nm $SiO_2$ on a silicon substrate in an argon-filled glovebox with both oxygen and water concentrations well below one part per million (1 ppm). The thickness of BP flake was determined by atomic force and high-resolution transmission electron microscopy measurements. The poly (methyl methacrylate) (PMMA) resist layer was spun on wafer and then patterned for metallization using a Vistec 100 kV electron-beam lithography system. Chromium/Gold (3/30 nm) were then evaporated and the following lift-off process in acetone formed the source, drain and voltage probe contacts on BP flakes. The 24-nm-thick $Al_2O_3$ top-gate dielectric was formed by atomic layer deposition (ALD) at 150 °C. Titanium/Platinum (1/10 nm) top-gate electrodes were fabricated using the same metallization procedure as discussed above. The top-gate electrode was designed to cover the entire transistor to ensure the full control of the channel.

### Conductance measurements using 4-probe scheme

All the electrical measurements were performed using an Agilent B1500A semiconductor parameter analyzer together with a Keithley 2612B source meter in a Lakeshore cryogenic probe station with 6 probe-arms. The static back-gate biases were applied using the source meter while the conductance as a function of top gate bias was measured using the Agilent parameter analyzer. The drain bias was set to be as low as 100 mV in order to minimize the drain-induced doping non-uniformity across the channel, thus ensuring the accurate determination of the conductance minima. The conductance of BP between probes 1 and 2 was obtained through dividing the channel current by the voltage drop between them. The impact of the contact resistance is eliminated by this 4-probe scheme.

### Acknowledgements

F.X. acknowledges the financial support through a Young Investigator Program (YIP) from the Office of Naval Research (N00014-15-1-2733). L.Y., H.W., and J.C. thank the support from the National Science Foundation EFRI-2DARE program (EFMA-1542815). Facilities use in Yale was partially supported by Yale Institute for Nanoscience and Quantum Engineering (YINQE) and NSF MRSEC DMR 1119826.

**Author contributions**

L.Y. and F.X. conceived and supervised the projects. B.D., H.J., H.T., Q.X., and H.W. participated in device fabrication. B.D., Q.G. and X.W. performed device characterizations. Y.X. and J.C. performed the device cross-section characterizations. V.T. and L.Y. did the modeling. F.X., L.Y., B.D., and V.T. drafted the manuscript. All the authors discussed the results and commented on the manuscript.

**Additional information**

Supplementary information is available in the online version of the paper. Reprints and permission information is available online at www.nature.com/reprints. Correspondence and requests for materials should be addressed to F.X. and L.Y.

**Competing financial interests**

The authors declare no competing financial interests.

**Figure Captions**

**Figure 1. Experimental scheme for black phosphorus bandgap tuning.** **(a)** The schematic view of the dual-gate BP thin film transistor for the realization of bandgap tuning. **(b)** The 4-nm-thick BP film conductance in linear (left axis) and logarithmic (right axis) scales as a function of the back gate bias ($V_{BG}$) at zero top gate bias ($V_{TG}$ = 0 V). The conductance $G_{12}$ is measured using a 4-probe scheme as shown in (a) to eliminate the effect of the contact resistance. Inset: an optical image of a typical dual-gate BP transistor. Scale bar: 10 μm.

**Figure 2: Bandgap tuning in a 4-nm-thick black phosphorus film.** **(a)** The 4-nm-thick BP film conductance as a function of top gate bias ($V_{TG}$) at different static back gate biases ($V_{BG}$) from -25 to 25 V. **(b)** The minimum conductance at the charge-neutrality point as a function of external displacement field. Inset: the top gate bias at which the minimum conductance occurs ($V_{TM}$) as a function of the back gate bias ($V_{BG}$). The linear relation indicates the effective doping compensation by the back and top biases, leading to an insulating state at the charge-neutrality.

**(c)** Solid lines: The calculated bandgap tuning properties for BP films consisting of 4, 5, 6, 7, 16, 17, 18, and 19 layers using a tight-binding model built upon the density functional theory (DFT). The BP films with different thicknesses exhibit distinctively different tuning properties. Red dots: The measured bandgap tuning results for the 4-nm thick black phosphorus film (~7 layers).

**Figure 3: Bandgap tuning in a 10-nm-thick black phosphorus film.** **(a)** The 10-nm-thick BP film conductance as a function of top gate bias ($V_{TG}$) at different static back gate biases ($V_{BG}$) from -25 to 25 V. **(b)** The minimum conductance at the charge-neutrality point as a function of external displacement field. Inset: the top gate bias at which the minimum conductance occurs ($V_{TM}$) as a function of the back gate bias ($V_{BG}$).

**Figure 4: Bandgap determination in 10-nm-thick black phosphorus using temperature-dependent measurements.** **(a)** The conductance of the 10-nm-thick BP film as a function of top gate bias ($V_{TG}$) at different temperatures from 120 to 295 K. The curves at three different back gate biases are plotted ($V_{BG}$ = 0, -15, and 15 V). The bandgap is measured through the temperature dependent minimum conductivity at the charge-neutrality point. **(b)** Conductance measurement curves similar to those in (a) at two different back gate biases ($V_{BG}$ = -25 and 25 V) **(c)** Dash (solid) lines: the calculated bandgap tuning properties for BP films consisting of 17, 19, and 21 layers using the tight-binding model without (with) additional bandgap dependent dielectric screening effect. Green dots: the measured bandgap tuning for a 10-nm-thick black phosphorus film (~19 layers).

# Figure 1: Experimental Scheme for Bandgap Tuning

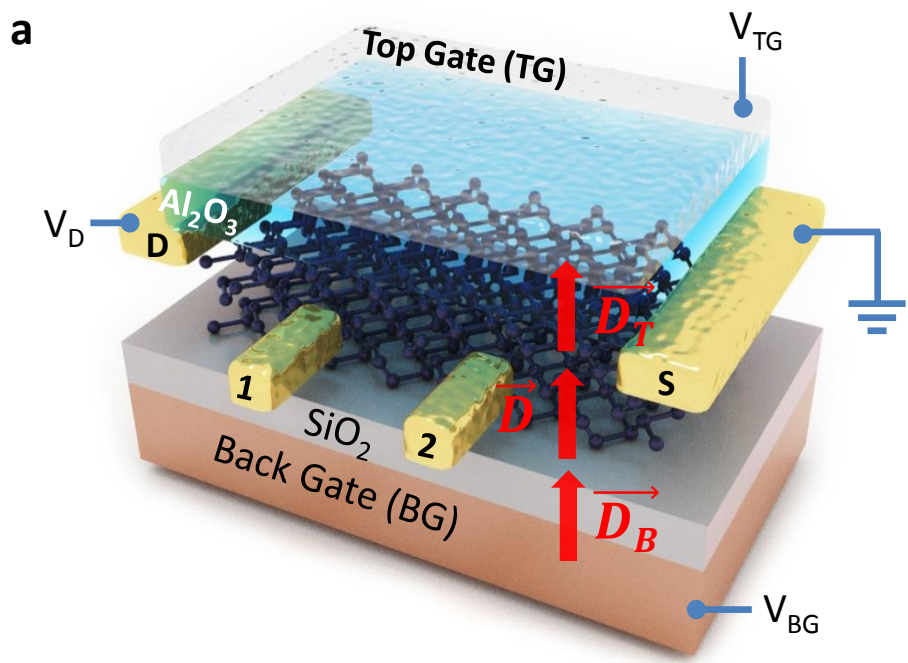

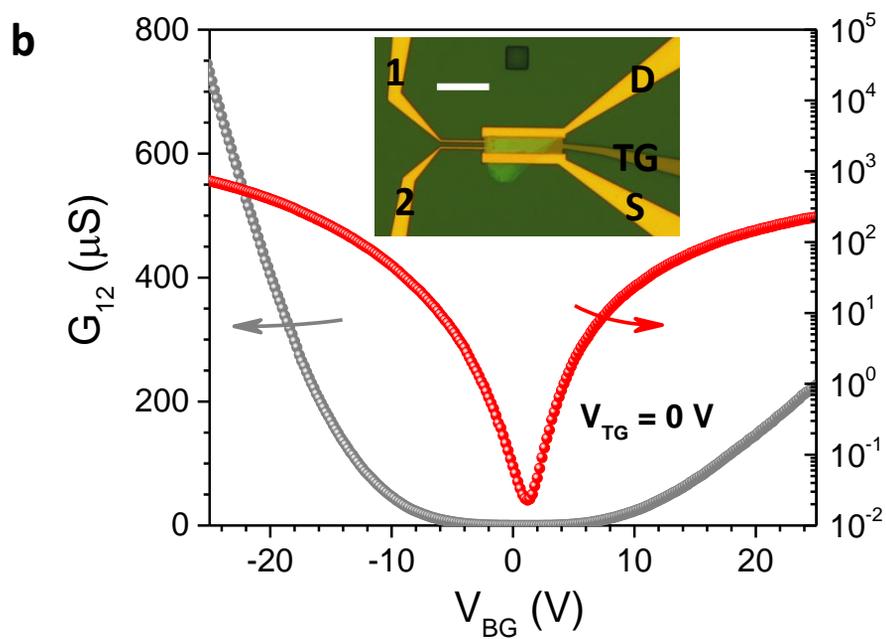

# Figure 2: Bandgap tuning in 4-nm-thick black phosphorus

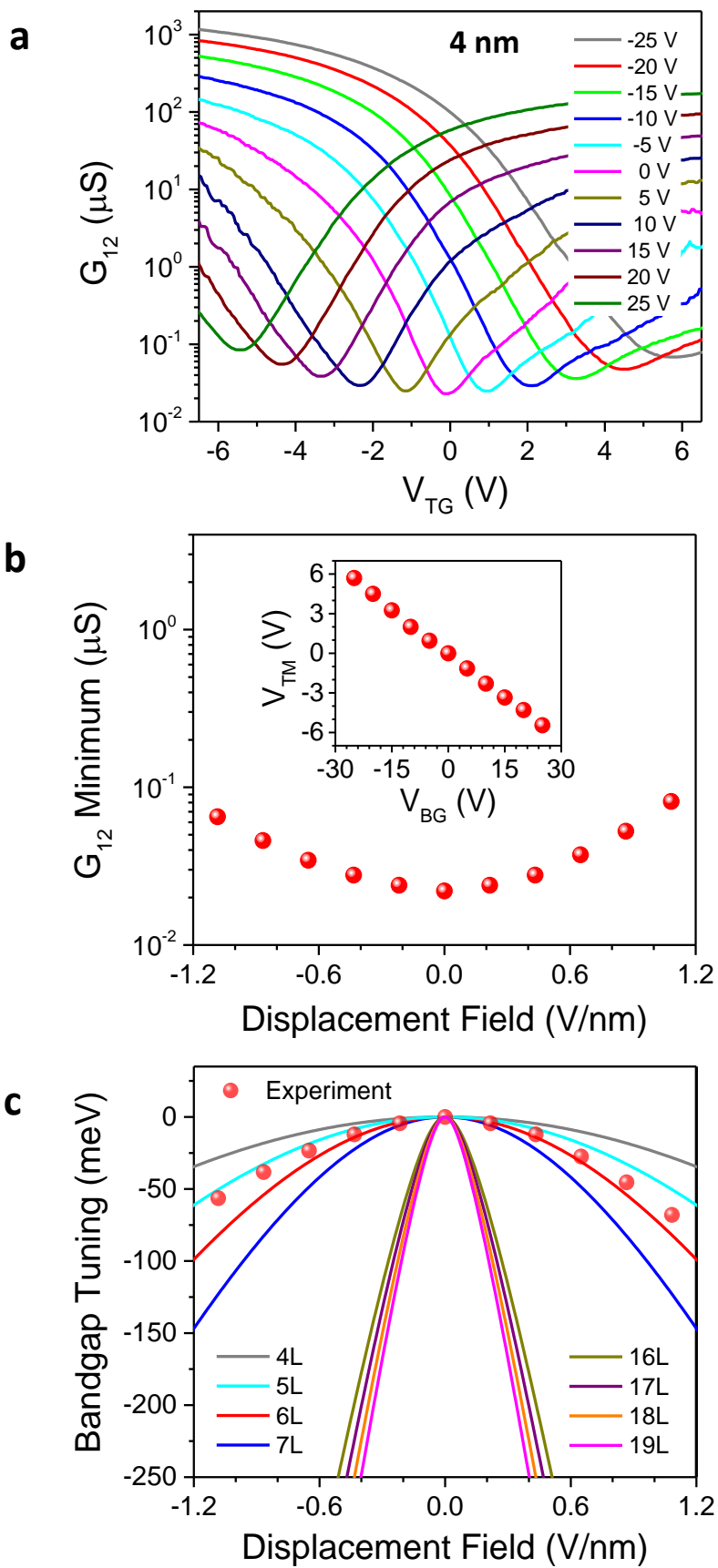

# Figure 3: Bandgap tuning in 10-nm-thick black phosphorus

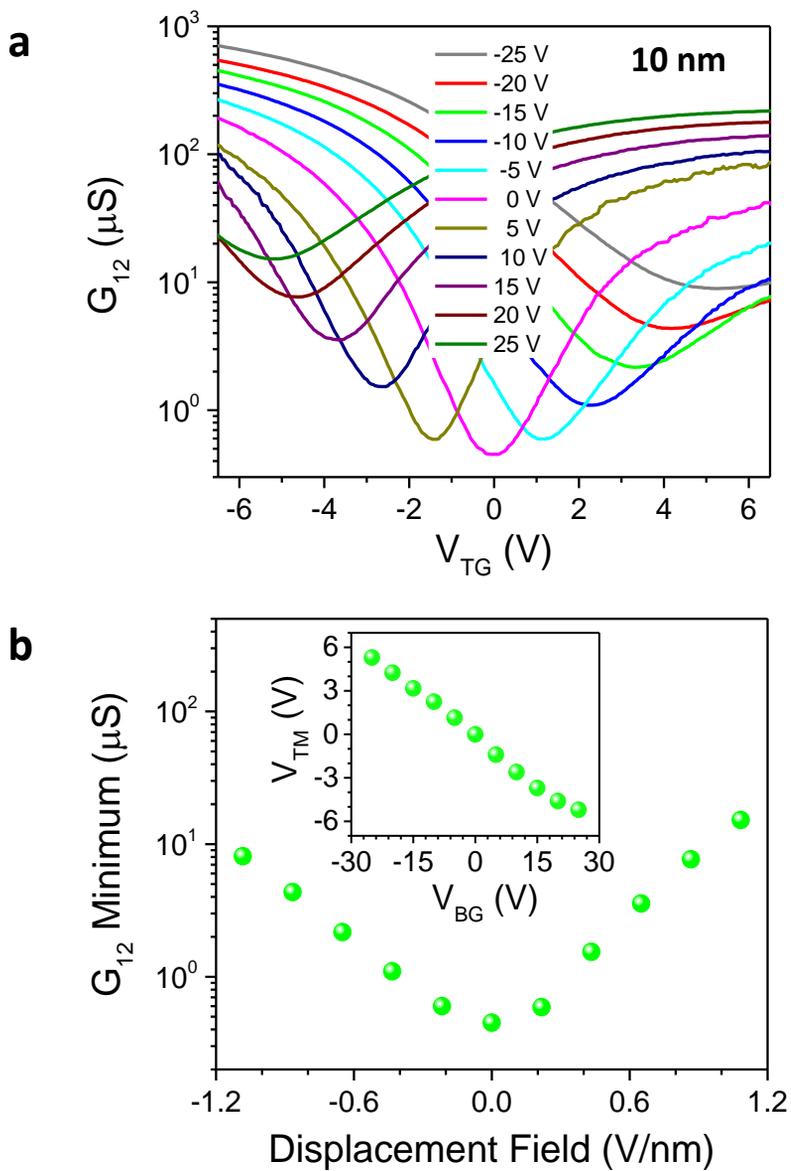

# Figure 4: Bandgap determination in 10-nm-thick black phosphorus using temperature-dependent measurements

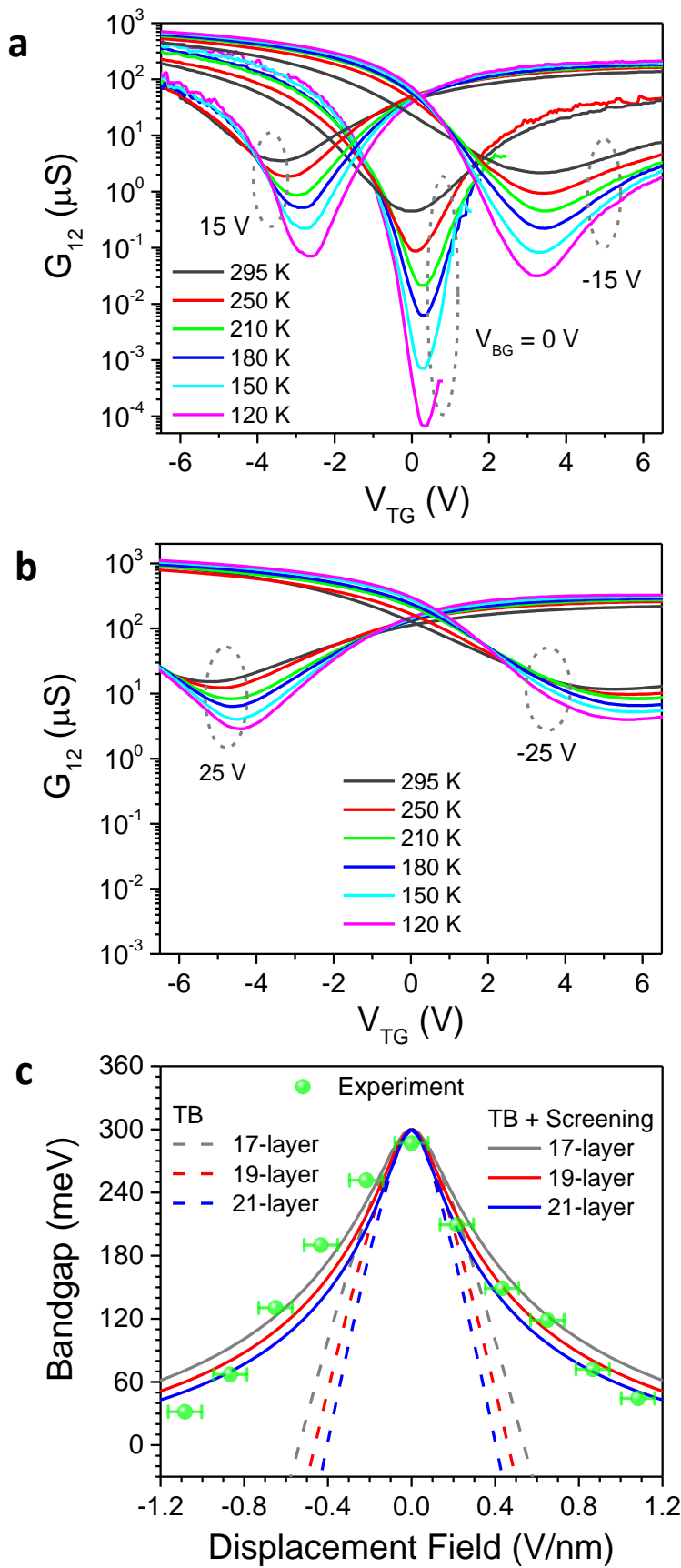

# Supplementary Information for "Efficient Electrical Control of Thin-Film Black Phosphorus Bandgap"

Bingchen Deng, Vy Tran, Hao Jiang, Cheng Li, Yujun (Terry) Xie, Qiushi Guo, Xiaomu Wang, He Tian, Han Wang, Judy J. Cha, Qiangfei Xia, Li Yang, and Fengnian Xia

## I. Thickness determination of black phosphorus thin film

We determine the thickness of the black phosphorus (BP) thin film using the standard atomic force microscope (AFM). Figure S1a shows a typical AFM image of device after the formation of the contact electrodes (source, drain contacts and voltage probes 1 and 2) and the atomic layer deposition of an $Al_2O_3$ gate dielectric. The measured thickness of this BP flake is about 6.5 nm, as shown by the line scan in Fig. S1b. Here the $Al_2O_3$ deposition does not affect the measurement accuracy of the BP flake thickness since it is deposited uniformly on both the substrate and the BP.

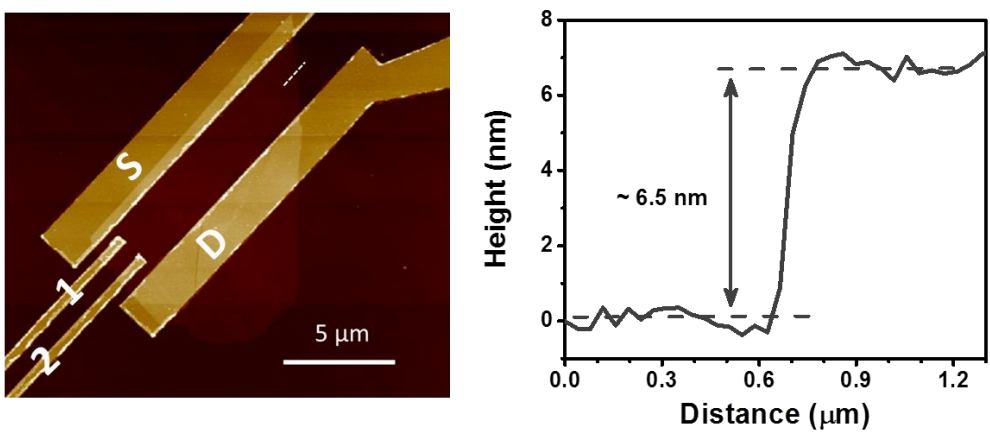

Figure S1. **(a)** An AFM image of the device after the formation of the contact electrodes and the top gate dielectric $Al_2O_3$. **(b)** A line scan performed along the white dashed line in **(a)**.

Similar to other elemental semiconductors such as silicon (Si) and germanium (Ge), it is well known that BP can oxidize. In order to accurately determine the BP thickness, separate from the possible surface oxide layer, we further characterized the sample using cross-section high-resolution transmission electron microscopy (HRTEM) to determine the intrinsic BP thickness. The results are summarized in Figure S2. It is rather clear that the phosphorus oxide (POx) at the top BP surface is 2 to 3 nanometers thick due to the exposure to external environment during processing. However, the bottom BP/$SiO_2$ interface is very sharp as shown in Fig. S2 and there is no obvious sign of oxidization because the sample exfoliation was performed in the glove box in which both the water and oxygen concentrations are well below one part per million (1 ppm). As a result, the intrinsic BP thickness in our devices is always 2 to 3 nm less than the value measured using AFM. In fact, our previous generations of BP devices have thicker phosphorus oxide on both top and bottom surfaces due to the exfoliation in ambient environment and lack of top gate dielectric protection. Hence, here the devices exhibit no sign of performance degradation up to a few months.

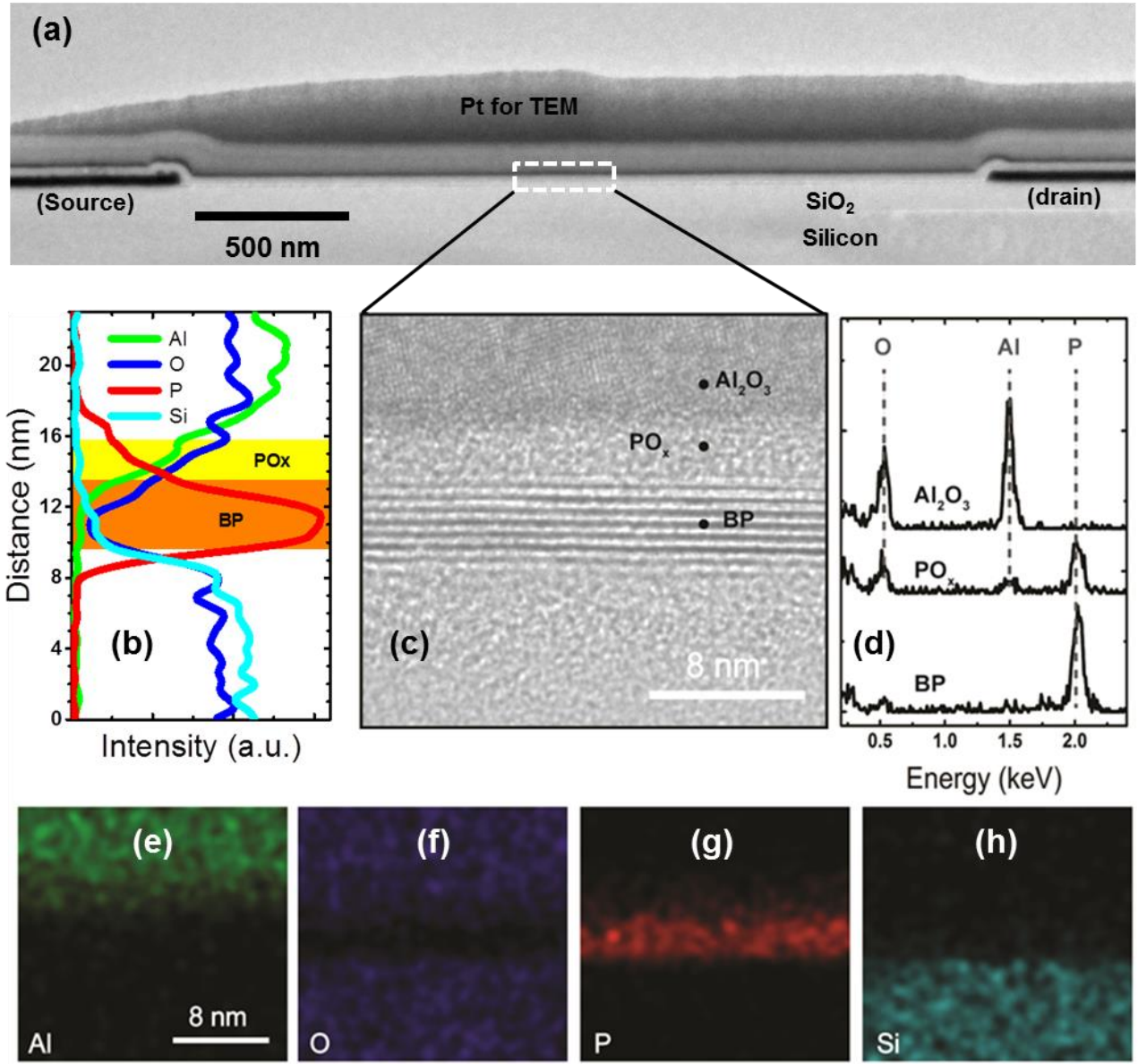

Figure S2. Cross-sectional HRTEM characterization of a BP device. **(a)** A cross-section TEM image showing the entire BP device. **(b)** Energy dispersive x-ray (EDX) spectroscopy line-scans showing the profiles of the aluminum (Al), oxygen (O), phosphorus (P), and silicon (Si) elements along the cross-section of the device. **(c)** High resolution cross-section TEM image showing the BP layer which is about 4-nm-thick. The POx layer between the BP layer and $Al_2O_3$ is 2 to 3 nm thick. **(d)** EDX spectra from three selected spots in (b). **(e-h)** Corresponding EDX elemental maps of Al (green), O (blue), P (red) and Si (cyan).

## II. Modeling of the BP bandgap under bias based on tight binding (TB) model

The bandgap reduction in multi-layer BP can be described by considering a tight-binding model which contains the energies of monolayer BP and coupling terms which describe the interlayer electronic state coupling. This model is motivated by examining the set of bands around the band gap for few-layer BP. Figures S3a and S3b show the band structure for 2-layer and 3-layer BP computed based on the density functional theory (DFT), respectively. It is clear that adding layers produces additional bands which look like the monolayer bands but are split off from each other due to interlayer coupling, like a particle tunneling between multiple wells.

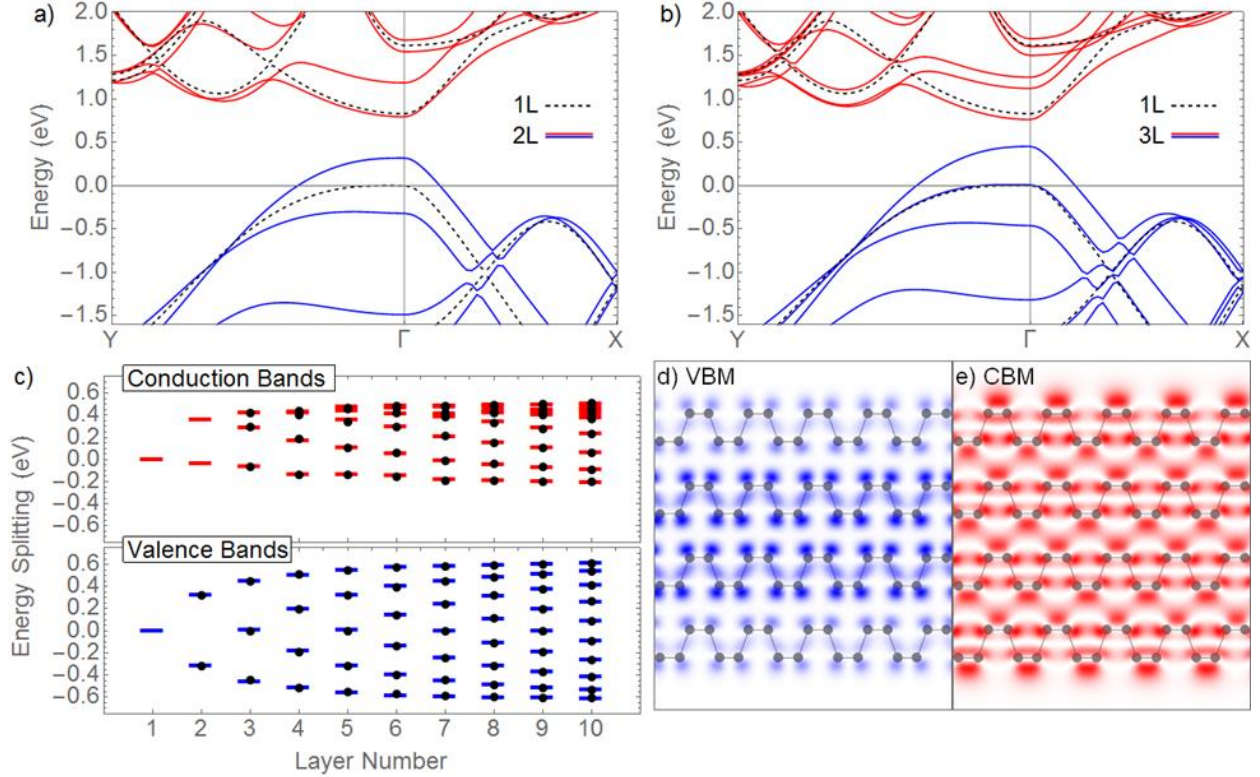

Figure S3. (a) The band structure of 2-layer BP compared with monolayer BP. (b) The band structure of 3-layer BP compared with monolayer BP. (c) The evolution of the energy of the valence band maximum (VBM) and conduction band minimum (CBM) states as the number of layers is increased. Colored lines represent direct DFT results and points are the fitted results using TB model. (d) and (e) The wave functions of the VBM and CBM states for 4-layer BP, respectively.

First, we discuss the model without an external electric field. The model Hamiltonian for n-layer BP can be written in matrix form as:

$$H_n = \begin{pmatrix} E_0 & \delta_1 & \delta_2 & \cdots \\ \delta_1 & E_0 & \delta_1 & \cdots \\ \delta_2 & \delta_1 & E_0 & \cdots \\ \vdots & \vdots & \vdots & \ddots \end{pmatrix}_{n \times n}$$

where the diagonal term $E_0$ is the energy of monolayer BP and the off-diagonal terms $\delta_i$ describe the coupling between layers. When the subscript $i=1$, $\delta_1$ is the nearest neighboring interlayer coupling; when the subscript $i=2$, $\delta_2$ is the second nearest neighboring interlayer coupling. In principle, the Hamiltonian is $k$ dependent, however since we are only interested in the valence band maximum (VBM) and conduction band minimum (CBM) states, we set $k = 0$, simplifying the model significantly.

The values of the parameters were obtained by fitting the model to DFT calculations. For n-layer BP, the original VBM/CBM is split into $n$ bands, which are fit to the $n$ parameters (the off-diagonal terms $\delta_i$) in the model. This direct fitting approach was performed for up to ten layers, and the comparison of the fitted model to DFT is shown in Fig. S3c. For the valence bands, the value of the nearest-neighbor term $\delta_1$, is 318 meV. The terms beyond $\delta_1$ were almost zero, and are treated as zero in the model. For the conduction bands, the coupling terms depend on the number of layers, but converge to a bulk limit, which is extrapolated and used for n > 10. For 10-layer to 20-layer BP, we assume that their local interlayer coupling is nearly the same as the bulk values. Thus we set $\delta_1^{cond} = 141$ meV and $\delta_2^{cond} = -85$ meV. For the conduction band the terms beyond $\delta_2$ were found to be small enough to ignore. The parameters for the conduction band are summarized in Table S1.

| Layers | 2 | 3 | 4 | 5 | 6 | 7 | 8 | 9 | 10 | Bulk |
|---|---|---|---|---|---|---|---|---|---|---|
| $\delta_1^{cond}$ | 196 | 170 | 176 | 172 | 163 | 165 | 163 | 161 | 159 | 141 |
| $\delta_2^{cond}$ |  | -76 | -82 | -73 | -89 | -93 | -94 | -95 | -97 | -85 |

Table S1. The parameters for the conduction band model Hamiltonian obtained by fitting to DFT calculations. All values are in meV.

To explain the layer dependence and additional coupling terms in the conduction bands we compare the VBM and CBM wave functions for 4-layer BP in Figs. S3d and S3e. It is clear that the VBM state is much more localized than the CBM state. The CBM state has considerable amplitude in the interlayer region, as well as extending from the surfaces. This leads to the second-neighbor coupling. Another feature of second-neighbor coupling is the asymmetric splitting of the CBM energies. From Fig. S3c we see that the valence bands are split symmetrically around the monolayer energy, while conduction band energy split is highly asymmetric. The single off-diagonal term in the valence band model controls the strength of the split, and additional terms in the conduction band model lead to asymmetric split.

Our model then contains effectively three parameters: $\delta_1^{val}$, $\delta_1^{cond}$ and $\delta_2^{cond}$, which are obtained by fitting to DFT calculations for up to 10 layers. This model describes the band gap evolution with respect to the layer number. To obtain the band gap in the presence of an out of plane electric field, an additional term is added, which shifts the potential of each layer due to the electric field. The models of conduction bands and valance bands can be written as:

$$H_n^c = \begin{pmatrix} E_0^c & \delta_1^{cond} & \delta_2^{cond} & \cdots \\ \delta_1^{cond} & E_0^c + \Delta & \delta_1^{cond} & \cdots \\ \delta_2^{cond} & \delta_1^{cond} & E_0^c + 2\Delta & \cdots \\ \vdots & \vdots & \vdots & \ddots \end{pmatrix}_{n \times n} \text{ and } H_n^v = \begin{pmatrix} E_0^v & \delta_1^{val} & 0 & \cdots \\ \delta_1^{val} & E_0^v + \Delta & \delta_1^{val} & \cdots \\ 0 & \delta_1^{val} & E_0^v + 2\Delta & \cdots \\ \vdots & \vdots & \vdots & \ddots \end{pmatrix}_{n \times n}$$

where $\Delta = \frac{E^{ext}}{\epsilon} * d$, $E^{ext}$ is the external electric field, $\epsilon = 8.3$ is the dielectric constant of intrinsic BP [S1], and $d = 0.53$ nm is the layer-layer distance. Here we emphasize that in our model $E^{ext}$ is the external field in vacuum. In our devices, the dielectric constant of the gate stack has to be taken into account. The potential on each layer is then shifted by an amount $\Delta$, which depends on the layer-layer distance and screened electric field in BP. Note that the addition of an electric field to the model requires no additional tunable parameters. Figure S4. shows the results obtained by tight binding model and DFT for 2-6 layers under bias; the tight binding model agrees with DFT extremely well. In the Fig. 2c of the main text, the calculated bandgap results of 4, 5, 6, 7, 16, 17, 18, and 19 layers of BP under bias using the tight binding model are plotted.

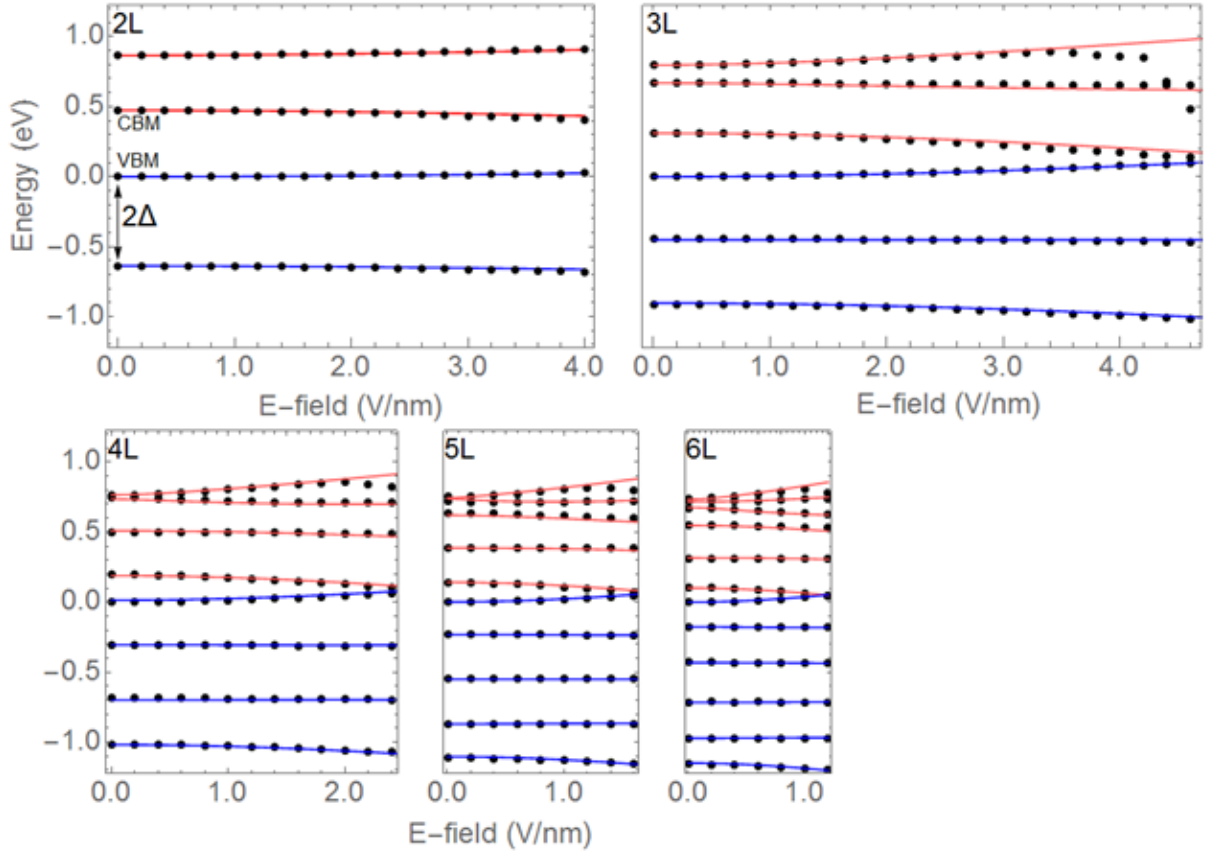

Figure S4. The energies of the VBM and CBM states for 2 to 6-layer BP as a function of applied electric field. The points are from DFT calculations, and the lines are from diagonalizing the model Hamiltonian. DFT and tight binding results agree well.

Although the tight model accurately reproduces the results of DFT calculations, it assumes the dielectric constant is the same for all layer numbers under all electric field conditions. Realistically, as the band gap decreases, the system should become more metallic, and the dielectric constant should increase. This effect is not significant for BP with a thickness of ~4 nm since the bandgap tuning is less than 100 meV and the original bandgap is estimated to be around 450 meV. However, for ~10-nm BP, this effect can be significant since the tuning exceeds 250 meV and the original bandgap is ~300 meV. To account for this, the dielectric constant in the expression for $\Delta$ is taken to be $\epsilon = \frac{E_0}{E_{gap}} \epsilon_{bulk}$, where $E_0$ is the intrinsic band gap under zero bias field [S2] and $\epsilon_{bulk} = 8.3$ is the original dielectric constant. The additional screening reduces the overall band gap reduction, as can be seen from Fig. 4c in the main text.

## III. Determination of the BP bandgap under bias

At each back-gate bias, we performed temperature-dependent measurements to determine the minimum conductance as a function of temperature. When the top gate bias reaches $V_{TM}$, the BP sample is at its overall intrinsic state and therefore we have:

$$\sigma_m = q n_i (\mu_e + \mu_h) \qquad (1)$$

where $\sigma_m$ is the minimum conductivity, $q$ is the elementary charge, $n_i$ is the intrinsic carrier density for both electrons and holes, and $\mu_e$ and $\mu_h$ are the mobility for electrons and holes, respectively. The minimum conductivity is a function of temperature primarily because we have [S3]:

$$n_i \propto T^{3/2} e^{-E_g/2k_B T} \qquad (2)$$

where $T$ is the temperature, $E_g$ is the material bandgap, and $k_B$ is the Boltzmann constant. In addition, both electron and hole mobility depend on the temperature. In BP thin films, previous experiments show that the carrier mobility has a temperature dependence of $T^{-1}$ [S4]. As a result, we have $\frac{\sigma_m}{T^{1/2}} \propto e^{-E_g/2k_B T}$. Therefore, the bandgap can be determined from the slope of the $\ln(\frac{\sigma_m}{T^{\frac{1}{2}}})$ v.s. $(\frac{1}{T})$ curves. A typical fitting curve at $V_{BG}$ of 15 V is plotted in Figure S5. At the high electric field, the bandgap can be very small. In this case, we replace the Eq. (2) above by the original Fermi-Dirac integral and performed the fitting again. The difference in the bandgap values at all biases is within 5%.

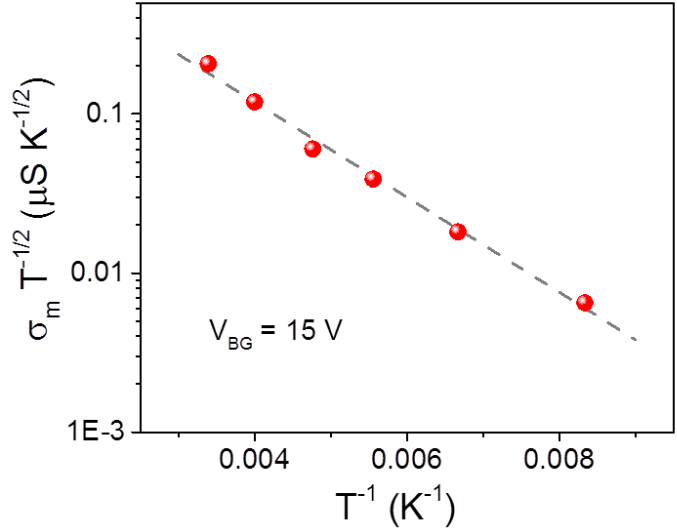

Figure S5. A typical fitting curve to determine the bandgap of the BP thin film under the bias ($V_{BG}$ = 15 V and $V_{TM} \approx$ -3 V). Red dots: measured results. Dash line: linear fitting curve.

## IV. A 23-nm-thick BP film under bias

We also performed similar conductance measurements on thicker BP thin films. Figure S6 denotes the conductance of a 23-nm-thick BP film under bias using the 4-probe scheme. Here the results are distinctively different from those in Figs. 2a and 3a in the main text. First, when the back gate bias is negative (-25 to -10 V), no obvious conductance minimum is observed. Second, when the back gate bias is positive (5 to 25 V), the $V_{TM}$, the top gate bias at which the conductance minimum occurs, is almost independent of the back gate bias. Both phenomena indicate that in thick BP thin-film, the electrostatic doping introduced by top and bottom gate biases cannot be compensated even if they are of the opposite polarity, because the free carriers screen the displacement field introduced by the gates. In this case, the BP channel can be approximately regarded as two rather independent channels controlled by the back and top gates, respectively. When the back gate bias is negative, holes accumulate in the bottom BP channel. Due to the high hole mobility, the bottom channel has a large conductance. The top gate mainly modulates the conductance of the top BP channel. Due to the relatively low electron mobility, the top BP channel conductance is low when $V_{TG}$ is positive, leading to the rather small change in overall conductance $G_{12}$ when $V_{TG}$ varies from 0 to 6 V. The overall $G_{12}$ conductance minimum can be observed when the bottom gate bias changes from 0 to 25 V. This is because at zero back gate bias the bottom BP channel conductance is minimized and at positive back gate bias the bottom BP channel conductance is relatively small due to the low electron mobility. Again, due to the free carrier screening effect, the negative top gate bias does not effectively compensate the electron doping in bottom BP channel, which leads to rather stable $V_{TM}$ when the back gate bias changes from 5 to 25 V.

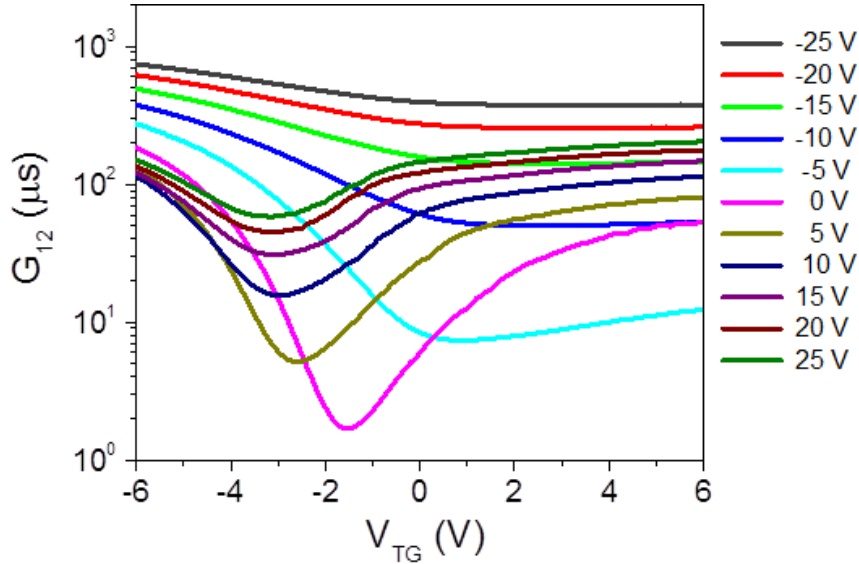

Figure S6. The overall conductance $G_{12}$ of a 23-nm-thick BP thin film as a function of top gate bias for fixed back gate bias from -25 to 25 V.